# ATTENTION-BASED GATED SCALING ADAPTATIVE ACOUSTIC MODEL FOR CTC-BASED SPEECH RECOGNITION

*Fenglin Ding, Wu Guo, Lirong Dai, Jun Du*

National Engineering Laboratory for Speech and Language Information Processing
University of Science and Technology of China, Hefei, China
flding@mail.ustc.edu.cn, {guowu,lrdai,jundu}@ustc.edu.cn

## ABSTRACT

In this paper, we propose a novel adaptive technique that uses an attention-based gated scaling (AGS) scheme to improve deep feature learning for connectionist temporal classification (CTC) acoustic modeling. In AGS, the outputs of each hidden layer of the main network are scaled by an auxiliary gate matrix extracted from the lower layer by using attention mechanisms. Furthermore, the auxiliary AGS layer and the main network are jointly trained without requiring second-pass model training or additional speaker information, such as speaker code. On the Mandarin AISHELL-1 datasets, the proposed AGS yields a 7.94% character error rate (CER). To the best of our knowledge, this result is the best recognition accuracy achieved on this dataset by using an end-to-end framework.

## 1. INTRODUCTION

With the extensive use of deep learning in automatic speech recognition (ASR), recognition accuracy has greatly improved over the past several years [1, 2]. However, ASR performance greatly deteriorates when there are mismatches between the training and the test data; these mismatches are caused by the different characteristics of acoustic variability, such as speaker, channel and environmental noises. An effective approach to dealing with these mismatches is acoustic model adaptation. This paper focuses on end-to-end (E2E) model adaptation using an attention-based gated scaling (AGS) scheme.

There are generally two classes of acoustic model adaptation techniques: feature and model space. A traditional feature-space technique is to transform the acoustic features to a normalized space, and then the adapted features are used to train the acoustic model. Maximum likelihood linear regression (MLLR) transforms and their feature-space variant (fMLLR transforms) [3, 4] are the two most widely used methods. For deep neural network (DNN)-based acoustic models, another effective method is to provide the network with auxiliary features that characterize speaker information to perform adaptation, such as i-vector [5, 6, 7] and speaker code [8, 9]. In model-space techniques, speaker-dependent (SD) parameters are estimated from a well-trained speaker-independent (SI) model using additional adaptation data. A straightforward idea is to retrain the parameters of the whole SI model. To avoid overfitting, regularization approaches, such as the Kullback-Leibler divergence (KLD) [10] and adversarial multitask learning (MTL) [11], were proposed. To prevent training large parameters, only small subsets of the network parameters for adaptation have been proposed [12, 13, 14]. In [15], P. Swietojanski *et al.* proposed learning hidden unit contribution (LHUC), in which the outputs of the SI hidden layer are reweighted by a set of SD parameters estimated from corresponding speaker adaptation data. LHUC or its variations [16, 17] can be viewed as a re-parameterizing adaptation with hidden scaling activations. All these methods yield recognition improvements on many ASR tasks.

Recently, researchers began training adaptive models on the fly instead of estimating the adaptive parameters from a well-trained SI model. Sequence summarizing neural network (SSNN) [18] and dynamic layer normalization (DLN) [19] are two typical techniques, where SD auxiliary networks are adopted to improve adaptive training and are jointly optimized with the main network. These methods greatly simplify model adaptation by using only one-pass training and not requiring additional adaptation data.

The aforementioned techniques have improved recognition accuracy by various degrees. However, all these works focus on a hidden Markov model (HMM) framework, and the adaptive mechanics are applied at the frame level. In this paper, we focus on end-to-end (E2E) systems and adopt the well-known connectionist temporal classification (CTC) [20] loss function for acoustic modeling. The CTC loss function is computed at the utterance level, and the adaptive training strategy is tailored for this property to achieve a better effect. We integrate the merits of LHUC and DLN and propose a novel attention-based gated scaling (AGS) scheme to reweight the outputs of the main network at the utterance level. To put more emphasis on the normalization of more character-discriminative deep features, we introduce an attention mechanism to enhance the scaling transformation in different frames. Further, AGS is implemented by auxiliary layers, which are jointly trained and optimized with the main network without additional adaptation data. We evaluate the proposed AGS algorithm on the AISHELL-1 corpus [21], an open-source Mandarin ASR task. Experimental results show that the proposed method greatly outperforms three typical types of adaptation algorithms proposed for conventional hybrid systems: LHUC, SSNNs, and DLN.

The rest of this paper is organized as follows: First, Section 2 provides a detailed description of our proposed acoustic model adaptation techniques. Section 3 describes the experimental setup and other details, including performance and comparisons with other adaptation methods, and Section 4 concludes the paper.

## 2. ATTENTION-BASED GATED SCALING

The proposed AGS framework is shown in Fig. 1. On the left is the conventional main network for acoustic modeling, which can be a feedforward neural network (FNN), a convolutional neural network (CNN) or a recurrent neural network (RNN); we adopt a Long Short-Term Memory (LSTM)-based RNN in this paper. On the right, the auxiliary AGS layers serve as a control network to fulfill the scaling transformation. The hidden layers of the main network are normalized by the corresponding AGS layers.

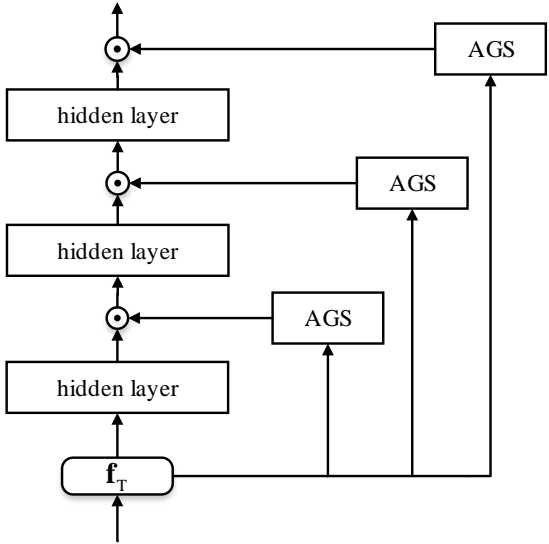

**Fig. 1.** The framework of the proposed attention-based gated scaling (AGS) adaptive network.

Suppose $\mathbf{h}_{l-1}$ and $\mathbf{S}_l$ denote the output of the layer and the outputs of the corresponding AGS layer, respectively, in Fig. 1; the output of the $l^{th}$ layer $\mathrm{D}_l(\mathbf{h}_{l-1})$ are rescaled as follows:

$$\mathbf{h}_l = \mathbf{S}_l \odot \mathrm{D}_l(\mathbf{h}_{l-1}). \quad (1)$$

where $\odot$ denotes a Hadamard product. The inputs of the AGS layers are the original or transformed acoustic features $\mathbf{f}_T$, where T denotes the frame length of the input utterance.

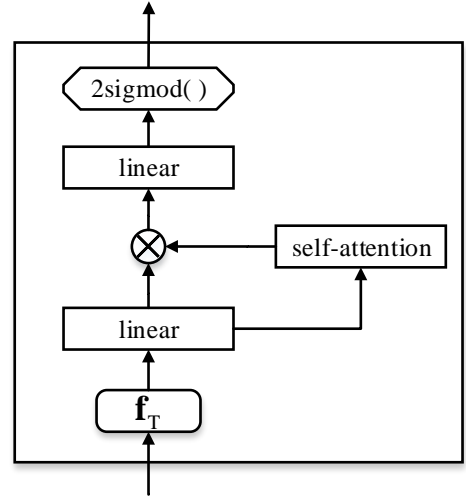

**Fig. 2.** The framework of the attention-based gated scaling (AGS) layer.

The self-attention [22] strategy is adopted in AGS to highlight the most relevant parts of the deep features for utterance-level acoustic modeling. Fig. 2 illustrates the structure of the proposed AGS layer. The attention weight $\boldsymbol{\alpha}$ for the whole utterance $\mathbf{f}_T$ can be calculated as:

$$\mathbf{K} = \mathbf{W}_k * \mathbf{f}_T, \quad (2)$$
$$\mathbf{Q} = \mathbf{W}_q * \mathbf{f}_T, \quad (3)$$
$$\mathbf{V} = \mathbf{W}_v * \mathbf{f}_T, \quad (4)$$
$$\boldsymbol{\alpha} = \mathrm{softmax}(\frac{\mathbf{K}^T \mathbf{Q}}{\sqrt{d_a}}). \quad (5)$$

where $\mathbf{W}_k$, $\mathbf{W}_q$ and $\mathbf{W}_v$ are $d_a \times d_f$ projection matrices; $d_a$ denotes the dimension of key: , query: and value; and $d_f$ denotes the dimension of the input feature $\mathbf{f}_T$.

The similarity scores between $\mathbf{K}$ and $\mathbf{Q}$ in Eq. 5 is directly computed in a dot-product attention mechanism for its speed and space efficiency.

The transformation matrix $\mathbf{S}_l$ is formed as a nonlinear activation of the reweighted features $\mathbf{C}$. A sigmoid function with amplitude 2 is used to constrain the range of the elements of the matrix to [0, 2], i.e.,

$$\mathbf{C} = \boldsymbol{\alpha} * \mathbf{V}, \quad (6)$$
$$\mathbf{S}_l = 2 * \mathrm{sigmoid}(\mathbf{W}_c^l \mathbf{C} + \mathbf{b}_c^l). \quad (7)$$

where $\mathbf{S}_l$ is a $\mathrm{T} \times d_h$ matrix and $d_h$ denotes the dimension of the hidden activations.

As the AGS auxiliary network operates on the whole utterance and all feature dimensions, the transformation matrix $\mathbf{S}_l$ can normalize the main network with both the time and feature dimension simultaneously. Furthermore, similar to the SSNN and DLN, the AGS auxiliary network is jointly trained and optimized with the main network on the fly, and the model adaptation is fulfilled by dynamically generating the scaling transformation parameters based on the input sequence rather than by learning them as other parameters in neural networks. This procedure makes the whole adaptation process simple and fast. Since a deeper network can achieve better recognition accuracy, different scaling transformation parameters are also generated for each layer of the main network, and this procedure can also improve performance.

## 3. EXPERIMENTS

### 3.1. Database

We evaluate the proposed methods on an open-source Mandarin speech corpus, AISHELL-1 [21]. All the speech files are sampled at 16 K Hz with 16 bits. The AISHELL-1 has 7,176 utterances from 20 speakers for evaluation (~10 hours). We use 120,098 utterances from 340 speakers (~150 hours) as the training set and 14,326 utterances from 40 speakers (~20 hours) as the development set. The speakers in the training, development and test set do not overlap.

### 3.2. Experimental setup

PyTorch toolkits [23] are used in our model training process. All the model parameters are randomly initialized and updated by Adam [24]. The acoustic feature is a 108-dimensional filter-bank feature (36 filter-bank features, delta coefficients, and delta-delta coefficients) with mean and variance normalization. According to statistical information obtained from the transcripts, there are 4,294 Chinese characters in the training set. Along with the added blanks, 4,295 modeling units are used in a grapheme-based CTC system. The trigram language model is used in the decoding procedure.

The network is trained to minimize the CTC loss function with an initial learning rate of 0.0001. The development set is used for learning rate scheduling and early stopping. We start to halve the learning rate when the relative improvement falls below 0.004, and the training ends if the relative improvement is lower than 0.0005; this latter decrease in improvement occurs usually at approximately 13 epochs. We also use a dropout rate of 0.3 for the LSTM layers to avoid overfitting.

### 3.3. Baseline, AGS and Contrastive Systems

The baseline acoustic modeling adopts a combination of CNNs and LSTM-based RNNs for good performance and high efficiency. For this baseline, the bottom two layers are 2D convolution layers with output channels of 64 and 256. Each convolution layer is followed by a max-pooling layer with a stride of 2 in the time dimension for finally down-sampling utterances to a quarter of their original length. After the CNN layers, there are three LSTM layers, each of which is a bidirectional LSTM layer with 512 units.

For the proposed AGS adaptive systems, the main components are the same as those of the baseline. Three auxiliary AGS layers are appended to the LSTM layer of the main network. The inputs of the AGS layers are the features down-sampled by the second CNN layer. The outputs of the AGS layers are fed to the corresponding LSTM layer of the main network, as shown in Fig. 1. We also use a dropout rate of 0.5 for the AGS layers to avoid overfitting.

We also build three types of adaptive models for comparison: an SSNN, an LHUC and a DLN model. For the SSNN systems, the SSNN is appended to the input of the baseline network. The configuration of the SSNN is similar to that proposed in [18], i.e., we use three fully connected layers of 512 units with hyperbolic tangent activation functions, except for the last layer, which has linear activation and 108 output units, the same number of units as for the input. A dropout rate of 0.1 is used for each layer of the SSNN. For the LHUC speaker adaptation systems, the trainable SD parameters are appended to the output of each LSTM layer and are only used for speaker adaptive training. A dropout rate of 0.3 is used for the SD parameters. For the DLN adaptation systems, the size of the utterance summarization feature vector is set to 256, and DLN is applied to all LSTM layers. A dropout rate of 0.5 is used for the utterance summarization feature extractor network.

### 3.4. Results

The character error rate (CER) is used as an evaluation criterion for our systems. We first investigate the effect of different configurations of the auxiliary AGS network on recognition accuracy. These configurations include different key and query dimensions $d_a$, and the head of multi-head self-attention of the AGS. The experimental results on the development sets of the different configurations are listed in Table 1.

As shown in Table 1, the minimum CER occurs when $d_a = 512$, but this result is only a marginal improvement over the CER obtained when $d_a = 256$ and the size of the parameters for the auxiliary network is almost doubled. Taking this into account, $d_a$ is set to 256 for the following experiments. By setting $d_a$ to 256, the AGS adaptation systems outperform the baseline systems on the development set with a 17.26% relative reduction of the CER and only a 6.3% relative increase in the size of the parameters. The AGS layers with multi-head attention are also explored; the number of heads is set to 2, 4, and 8, and the key/query dimension for each head is 256 divided by the number of heads. As shown in Table 1, multi-head AGS does not further reduce the CER. Therefore, single-head attention is used in the following experiments.

| $d_a$ | Head | Parameter Size (M) | CER (%) |
|---|---|---|---|
| baseline | - | 85.82 | 8.46 |
| 256 | 1 | 91.22 | 7.00 |
| 512 | 1 | 96.61 | 6.88 |
| 1024 | 1 | 107.38 | 7.12 |
| 256 | 2 | 91.22 | 7.15 |
| 256 | 4 | 91.22 | 7.05 |
| 256 | 8 | 91.22 | 7.25 |

**Table 1.** The ASR CER (%) of the AGS system with all adapted LSTM layers for different configurations of the auxiliary AGS network on development set of AISHELL-1.

In the following experiments, we investigate how many and which hidden layers should be adapted. The combinations of different adaptation layers are investigated. The results on the test set are summarized in Table 2. The baseline (no adaptation) performs well with a CER of 8.46% on the development set and 9.96% on the test set. If only one LSTM layer is adapted, the higher layer (3$^{rd}$ layer) achieves better performance than that of the lower layer (1$^{st}$ layer): on the development set, the CERs for the higher and lower layer are 7.60% and 8.34%, respectively; on the test set, the CERs for the higher and lower layer are 8.68% and 9.46%, respectively. Furthermore, the CER steadily decreases as the number of adapted layers increases. When all three LSTM layers are adapted, the proposed AGS algorithm achieves the best CER of 7.94% on the test set; this error rate is a reduction of 20.28% compared to the CER of the baseline system and is, as far as we know, the best performance on this dataset using the E2E ASR systems.

| Adapted LSTM layer | Dev. set | Test set |
|---|---|---|
| Baseline | 8.46 | 9.96 |
| 1 | 8.34 | 9.46 |
| 2 | 7.92 | 9.25 |
| 3 | 7.60 | 8.68 |
| 1,2 | 7.81 | 8.93 |
| 2,3 | 7.35 | 8.46 |
| 1,2,3 | **7.00** | **7.94** |

**Table 2.** The ASR CER (%) of the AGS adaptation system with different adapted LSTM layers on development and test set of AISHELL-1.

The experimental results of all the systems are listed in Table 3. For the other three contrastive adaptive training systems, all three LSTM layers of the main networks are adapted, and this configuration obtains the best performance. All the adaptive training algorithms also obtain CER reduction to some extent. The LHUC algorithm performs the best among the three contrastive systems, but the proposed AGS performs even better than the LHUC algorithm.

| System | Adapted LSTM layer | Dev. set | Test set |
|---|---|---|---|
| Baseline | - | 8.46 | 9.96 |
| SSNN | 1,2,3 | 8.30 (1.89) | 9.71 (2.51) |
| LHUC | 1,2,3 | 7.67 (9.34) | 8.78 (11.85) |
| DLN | 1,2,3 | 7.82 (7.57) | 8.92 (10.44) |
| AGS | 1,2,3 | **7.00 (17.26)** | **7.94 (20.28)** |

**Table 3.** The ASR CERs (%) of different algorithms on development and test set of AISHELL-1. Relative improvements are given in parentheses.

## 4. CONCLUSIONS

In this paper, we propose a novel adaptive training technique for E2E CTC acoustic modeling. Unlike previous works, our method normalizes the hidden features along both the time and feature dimensions simultaneously through a rescaling transformation augmented by an attention mechanism. Moreover, the adaptive training does not depend on any adaptation data or embedding of acoustic variability, thereby making it easier to implement on an E2E system and deal with different kinds of mismatched acoustic conditions. Experiments conducted on the AISHELL-1 dataset show that our proposed methods greatly improve the recognition accuracy of the baseline system. Further, our adaptation scheme also outperforms three typical adaptation algorithms proposed for conventional hybrid systems under the same test conditions.

## 5. ACKNOWLEDGEMENTS

This work was partially funded by the National Key Research and Development Program of China (Grant No. 2016YFB100 1303) and the National Natural Science Foundation of China (Grant No. U1836219).